\documentclass[preprint,prc,aps,floatfix,showpacs,showkeys]{revtex4}
\usepackage{epsfig}
\usepackage {graphicx}
\begin{document}
\title{Inclusive Spectra of (p,xp) and (p,xd) Reactions on
$^{90,92}$Zr and $^{92}$Mo Nuclei at E$_{p}$=30.3 MeV}
\author{A. Duisebayev, K. M. Ismailov}
\affiliation{Institute of Nuclear Physics, National Nuclear
Center, Republic of Kazakhstan}
\author{I. Boztosun}
\affiliation{Department of Physics, Erciyes University, Kayseri,
Turkey}
\date{\today}
\begin{abstract}
New experimental data for the inclusive reactions (p,xp) and
(p,xd) on isotopes of the nuclei $^{90,92}$Zr and $^{92}$Mo, have
been measured at E$_{p}$=30.3 MeV, which has not been investigated
in detail so far. We show the extension of the pre-equilibrium
reactions to this energy region and interpret the results of these
experiments. Moreover, we display the mechanism of the reaction
and the level of energy-dependence. The adequacy of the
theoretical models in explaining the measured experimental data is
also discussed. In our theoretical analysis, the contributions of
multi-step direct and compound processes in the formation of
cross-sections are determined and we assert that the traditional
frameworks are valid for the description of the experimental data.
\end{abstract}
\pacs{25.40.-h, 24.60.Dr, 24.60.Gv, 24.50.+g}
\keywords{pre-equilibrium reactions, (p,xp) and (p,xd) reactions
on $^{90,92}$Zr and $^{92}$Mo, Hauser-Feshbach model, multi-step
direct and compound processes} \maketitle

\section{INTRODUCTION}
Working out the pre-equilibrium decay mechanism in nuclear
reactions, which reflects the dynamics of the formation of the
excited system and its evolution to the equilibrium state, remains
an actual problem of the nuclear reaction theory. The problem is
largely connected with obtaining the new experimental data on
double-differential cross-sections in (p,xp), (p,xd) {\it etc.}
reactions with different proton energies. These reactions play a
role in the applied researches on secure and wasteless nuclear
power system creation (accelerator+subcritical reactor). In this
respect, there is a problem of determining the spatial and power
distribution of the secondary particles, generated not only during
the transition of the primary proton beam of target assembly and
neutron flow, but also of more composite ($^{2,3}$H, $^{3,4}$He)
particles, which can represent themselves as initiators of the
reactions by emitting neutrons.

In this paper, the zirconium and molybdenum elements have been
chosen as the objects of our investigation since they are the
construction elements of a hybrid nuclear energy plant.

Actually, in the past, there have been experimental studies of the
(p,xp) data on $^{98}$Mo at incident energies around 26 MeV by
Watanabe {\it et al.} \cite{ref1}. This group has measured the
energy spectra of emitted protons between angles 30$^{\circ}$ to
150$^{\circ}$, at intervals of 10$^{\circ}$. But, there are no
experimental data at outgoing proton energies around 7 MeV or less
because the protons have been stopped completely in the silicon
surface-barrier $\Delta $E detector, which was 300 micron thick.
They have analyzed the proton-induced reactions theoretically by
using the FKK-GNASH code based on the FKK theory. The experimental
nucleon emission spectra at 26 MeV have been reproduced well by
the calculations including pre-equilibrium MSD and MSC emission,
direct collective excitation to low-lying discrete levels, and
Hauser-Feshbach equilibrium emission, in a quantum-mechanical way.

There are also the experimental (p,xp) spectra for $^{98}$Mo at
12, 14 and 16 MeV \cite{ref2,ref3} and for $^{90}$Zr, $^{54,56}$Fe
and $^{93}$Nb at 26 MeV \cite{ref4} from the Kyushu group. In the
first case, the target was the self-supporting metallic foil
$^{98}$Mo, with a thickness of 450 $\mu $g/cm$^{2}$. Emitted
protons were detected with a $\Delta $E-E counter telescope
consisting of 75 $\mu $m and 2000 $\mu $m silicon surface
detectors. Proton energy spectra were measured at intervals of
10$^{\circ}$, from 30$^{\circ}$ to 160$^{\circ}$. The (p,xp)
spectra were analyzed on the basis of the exciton model and the
Hauser-Feshbach model, in which isospin conservation was taken
into account.

In the second case, they used self-supporting metal foils of $^{54
- 56}$Fe, $^{90}$Zr and $^{93}$Nb with a thickness of 500 $\mu
$g/cm$^{2}$, at 26 MeV. The thicknesses of $\Delta $E and E
silicon detectors were 30, 200 and 5000 $\mu $m. The energy
spectra were measured in the angle range 30-150$^{\circ}$, at
intervals of 10$^{\circ}$. These data were analyzed pursuant to
the FKK theory for the pre-equilibrium process by using the code
FKK-GNASH and pursuant to the Hauser-Feshbach model for the
compound process. The calculated results showed a good agreement
with the experimental spectra for both cases.

In our experiment, we focus on investigating two issues: Firstly,
there are no experimental data at outgoing proton energies around
7 MeV or less since, from Watanabe {\it et al.}'s measurement, the
protons have been stopped completely in the silicon
surface-barrier $\Delta $E detector, which was 300 micron thick.
In order to overcome this problem and to cover the soft part of
the spectra, we have used a 50 micron $\Delta $E-detector,
therefore, the energy range is started from 2.5 MeV for our
measurements. Secondly, the protons in the energy region of 30 MeV
have not been studied in detail \cite{ref1,ref2,ref3,ref4} and
extending the experiment in this direction allows us to view the
mechanisms of the reaction and the level of energy-dependence in
detail and to use these observations for adequate analysis within
the framework of the FKK theory. Undoubtedly, the extraction of
experimental information on channels of reactions with the
emission of complex particles (deuteron, tritium, $^{3}$He and
$^{4}$He) remains interesting.

In the next section, we present our experimental method, the
details of the measurement and the experimental results. Section
\ref{analysis} is devoted to the theoretical analysis of the
measured experimental data by the exciton model and quantum
mechanical representations. Finally, Section \ref{conc} gives our
summary and conclusion.

\section{EXPERIMENT AND RESULTS}
\label{exp}

The inclusive cross-section measurements of the reactions (p,xp),
(p,xd), originated by protons with E=30.3 MeV, on isotopes of the
nuclei $^{90, 92}$Zr and $^{92}$Mo, have been carried out on the
cyclotron U-150M of the Institute of Nuclear Physics NNC Republic
of Kazakhstan, within the range of 15-150$^{\circ}$, at intervals
of 15$^{\circ}$. Typically, intensities between 20 and 40 nA have
been utilized with a beam energy resolution of 0.3{\%}.

For the registration and identification of reaction products in
the whole energy range, a system has been designed with a
three-detection telescope ($\Delta $E$_{1}-\Delta
$E$_{2}$-E$_{3})$, consisting of two silicon surface-barrier ORTEC
detectors and a scintillation detector CsI (Tl) with a total
absorption of E$_{3}$. The thicknesses of detectors were $\Delta
$E$_{1}$=50 $\mu $m, $\Delta E_{2}$=300 $\mu $m and E$_{3}$=25000
$\mu $m correspondingly. The solid angle subtended by a telescope
of detectors was equal to $\Omega $=2.9 10$^{-5}$ sr. The spectra
of protons have been registered from a threshold, which is defined
by absorption in the first detector $\Delta $E$_{1}$, and by
maximum energy absorption in the second detector $\Delta $E$_{2}$
in an interval, and these spectra have been identified by a matrix
of coincidence ($\Delta $E$_{1}$x$\Delta $E$_{2})$. The second
interval of registration has been determined by energy of
fragments in the detector $\Delta $E$_{2}$, down to the energy of
total absorption in the detector E$_{3}$, at which a matrix of
identification ($\Delta $E$_{2}$xE$_{3})$ corresponds. Thus, the
power spectrum of protons in the studied reactions has been
measured within an energy range starting from a threshold of
E$_{p} \approx $2.5 MeV up to $ \approx $30 MeV, and deuterons $
\approx $ 2.5$\sim $20 MeV.

\begin{figure}
\centerline{\includegraphics{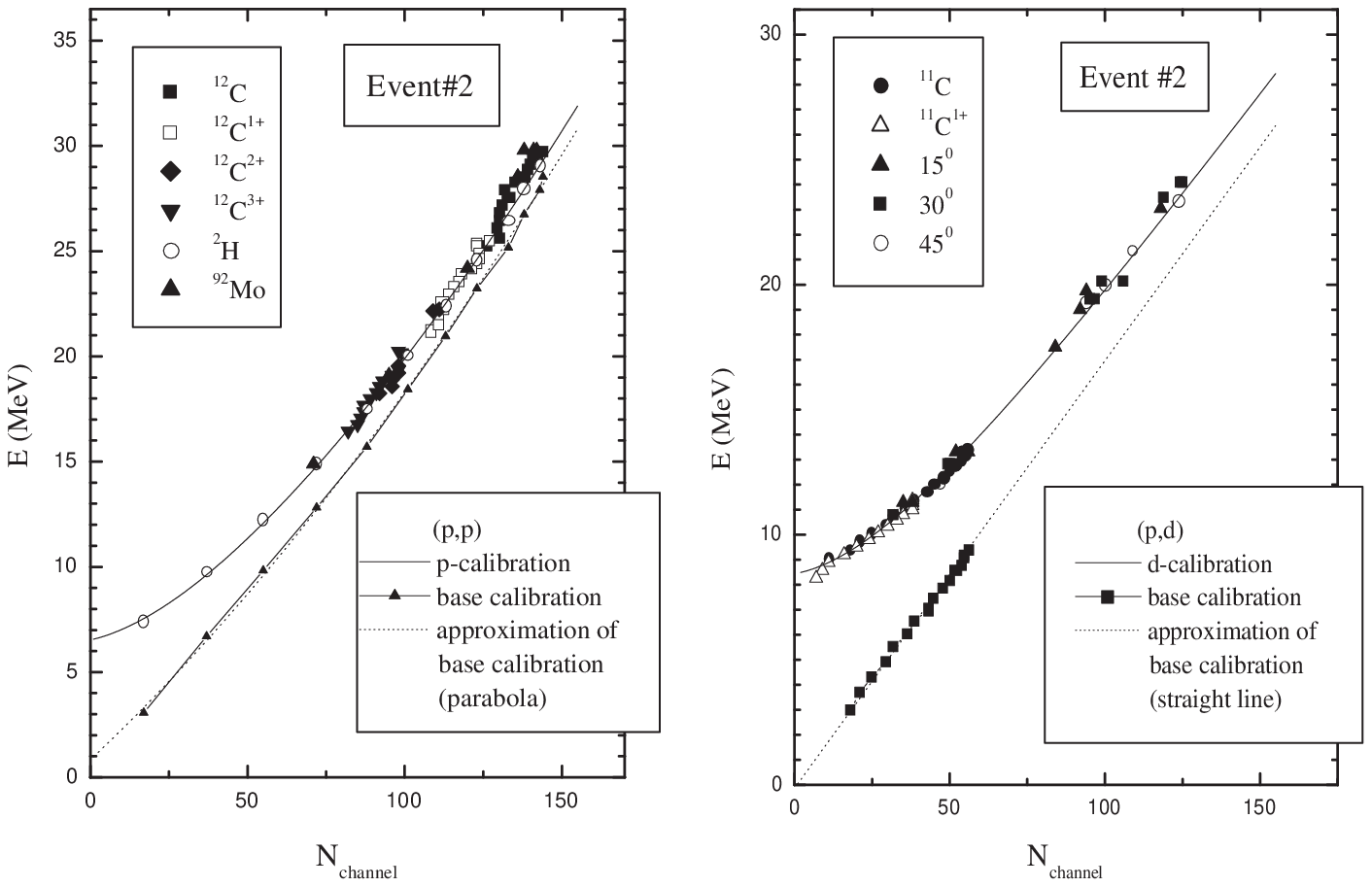}} \vskip-13.0cm
\caption{Calibration characteristics of the telescope $\Delta
$E$_{2}$xE$_{3}$ for protons and deuterons} \label{fig1}
\end{figure}

The energy calibration of a spectrometer has been carried out on
kinematics of levels of residual nuclei in the reaction $^{12}$C
(p,x) and protons of recoil. The base calibration $\Delta
$E$_{1}$x$\Delta $E$_{2}$ is approximated by a straight line and
it does not depend on any kind of fragments, whereas for events
$\Delta $E$_{2}$xE$_{3}$ (detector CsI (Tl)), the base calibration
represents a parabola for protons and a straight line for
deuterons. This is shown in Figure \ref{fig1}, where the base
calibration expresses the relation between the channel number in
line spectra and the lost energy in the detector. It is found by
subtracting the particle energy, lost in the target and the
detectors, from the kinetic energy. The energy of the particle
before hitting into the telescope of detectors is determined by
using this base calibration and by restoring the losses in the
detector. After that, we define the energy of the emitted particle
by adding the losses in the target. The full energy resolution of
the spectrometer has amounted to 800 keV for $\Delta
$E$_{2}$xE$_{3}$ and $ \approx $400 keV for $\Delta
$E$_{1}$x$\Delta $E$_{2}$.

\begin{figure}
\centerline{\includegraphics[width=4.86in,height=3.06in]{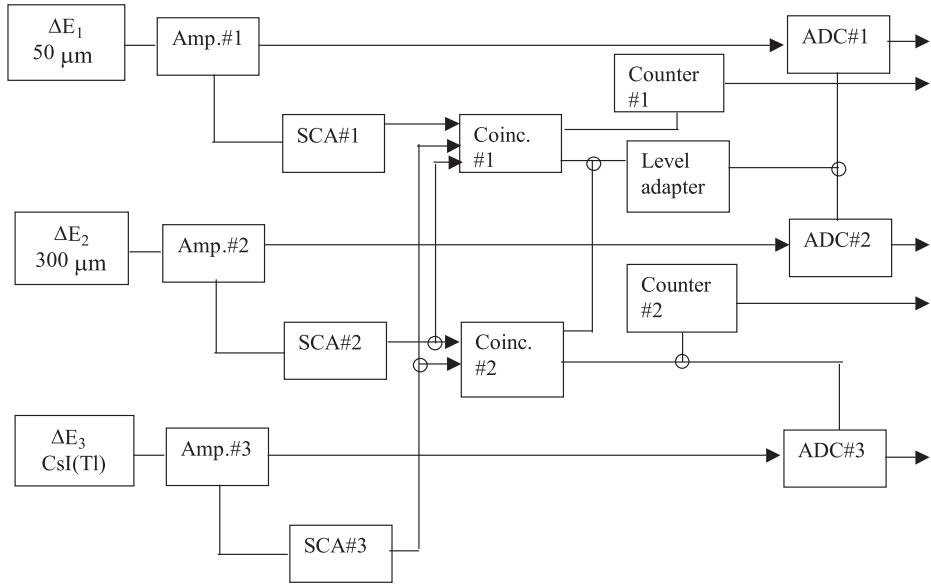}}
\caption{The block-scheme of the registration system:
Amp.{\#}1,2,3 - Spectroscopic Amplifier; SCA{\#}1,2,3 - Single
Channel Analyzer; Coinc.{\#}1,2 - Scheme of Coincidences;
Counter{\#}1,2 - Counter Scheme; ADC{\#}1,2,3 - Analog-Digital
Converter, Level Adapter - impedance matcher of levels.}
\label{fig2}
\end{figure}

The electronic scheme of the measurement is presented in Figure
\ref{fig2}, the main bends of which are four-entrance
spectroscopic amplifiers, multi-port (up to eight) analog-digital
converters, discriminators and pulse counters. The main complexity
during the set-up of the system is connected with the installation
of a threshold of the discriminator SCA{\#}3 for avoiding an
overload by false coincidences with gamma-quantums, that leads to
the formation of a {\it slot} between events $\Delta
$E$_{1}$x$\Delta $E$_{2}$ and $\Delta $E$_{2}$xE$_{3}$, in the
width of 2 MeV for protons and about 3 MeV for deuterons. The dead
time of the system makes from 7 up to 1 {\%} and is controlled by
the Counters 1-2.

In the experiments, the self-supporting targets of isotopes
$^{90,92}$Zr and $^{92}$Mo made by a method of electrochemical
evaporation have been used, the characteristics the one of which
is listed in Table \ref{tab2}.

\begin{table}
\caption{Characteristics of target nuclei} \label{tab2}
\begin{center}
\begin{tabular}{llll}\hline\hline
Isotope & $^{90}$Zr & $^{92}$Zr & $^{92}$Mo  \\\hline
Thickness, mg/sm$^{2}$ & 2.13 & 0.8 & 0.51  \\
Enrichment, {\%} & 95 & 97 & 95  \\
\hline\hline
\end{tabular}
\end{center}
\end{table}

In Figures \ref{fig3} and \ref{fig4}, the spectra of
double-differential cross-sections (d$^{2}\sigma $/dEd$\Omega )$
are shown. The total cross-sections of (p,xp) and (p,xd)
reactions, based on the integral (d$\sigma $/dE), are listed in
Table 2.

\begin{table}
\caption{Experimental partial cross-sections (mb) of reactions
(p,xp) and (p,xd)}\label{tab3}
\begin{center}
\begin{tabular}{lllll}\hline\hline
Reaction & OER\footnote{OER denotes Outgoing Energy Range} (MeV) &
$^{90}$Zr & $^{92}$Zr & $^{92}$Mo \\ \hline
(p,xp) & 3-27 & 973.3 $\pm $ 8.5 & 839.6 $\pm $ 6.4 & 1229.0 $\pm $ 10.2 \\
(p,xd) & 4-17 & 47.39 $\pm $ 0.59 & 28.83 $\pm $ 0.38 & 33.25 $\pm $ 0.51 \\
\hline\hline
\end{tabular}
\end{center}
\end{table}

In Figure \ref{fig5}, the experimental cross-sections of the
studied reactions are shown. Moreover, three areas connected with
different decay mechanisms for a channel with emitting protons are
observed. The first one is the area of equilibrium components of a
cross-section with a maximum at E$_{p} \approx $5 MeV, the
pre-equilibrium component (E$_{p} \approx $12-20 MeV) and the
range of E$_{p}>$20 MeV, conditioned by direct mechanisms
following it.

For a channel with emission of deuterons (p,xd), the situation is
less determined. Typically for this channel, the cross-section of
the equilibrium components is sharply overwhelmed (E$_{d} \approx
$5 MeV). It is remarkable that the form of deuteron spectra is
different for zirconium isotopes and $^{92}$Mo nuclei.

\section{ANALYSIS AND RESULTS}
\label{analysis}

There have been many theoretical approaches used to describe the
equilibrium reactions data over a wide range of incident energies
(see references \cite{Feshbach,Hodgson,Bonetti1,Bonetti2} for a
detailed discussion). In this paper, we analyze the experimental
cross-sections data of all the reactions pursuant to both the
Hauser-Feshbach theory -while taking into account the
multi-particle emission of both single-shot (protons, neutrons)
and two-charging fragments (deuterons, $\alpha $-particles)-, and
the stringent quantum-mechanical theory (program EMPIRE II
\cite{ref5}). Thus, the contributions of statistical direct and
compound processes for the reactions (p,xp) and (p,xd) have also
been calculated.

The approach to statistical multi-step direct reactions is based
on the multi-step direct theory of pre-equilibrium scattering to
the continuum, originally proposed by Tamura, Udagawa and Lenske
\cite{ref6}. Since then, the approach has been revised especially
in the part related to the statistical and dynamical treatment of
nuclear structure.

The evolution of the projectile-target system from small to large energy
losses in the open channel space is described in the MSD theory with a
combination of direct reaction (DR), microscopic nuclear structure and
statistical methods. As typical for the DR-approach, it is assumed that the
closed channel space, {\it i.e.} the MSC contributions, have been projected out
and can be treated separately within the multi-step compound mechanism.

The modelling of multi-step compound processes follows the
approach of Nishioka {\it et al.} (NVWY) \cite{ref7}. Like most of
the pre-compound models, the NVWY theory describes the
equilibration of the composite nucleus as a series of transitions
along the chain of classes of closed channels in increasing
complexity. In the present context, we define the classes in terms
of the number of excited particle-hole pairs (n) plus the incoming
nucleon, {\it i.e.} excitons. Thus, the exciton number is N = 2n+1
for nucleon induced reactions. Assuming that the residual
interaction is a two-body force, only neighbouring classes are
coupled ($\Delta $n = $\pm $1).

In all cases, the parameter of level density had already been
determined by Gilbert-Cameron parameterization \cite{ref8}. For
comparison, the calculation of the inclusive integral
cross-section with the parameter of level density a=A/8 has been
carried out, where A is the nuclear mass number.

The optical potential of Becchetti-Greenlees \cite{Becchetti} for
proton and neutron channels  and the optical potential of
Perey-Perey \cite{Perey} for deuteron channel have been used in
calculations of reaction transmission coefficients.

The results of these calculations are shown in Figure \ref{fig6}.
It is shown that the contribution of multi-particle compound of
the mechanism determines the emission of protons from 2.5 MeV up
to 10 MeV, and the contribution of the multi-step direct process
extends from 5 MeV up to the kinematic threshold. The shape of
(p,xp) reaction integral spectra is determined by the multi-step
direct processes. At the same time, the shape of spectra within
the indicated theory that can be connected to the contribution of
single-step direct mechanisms, which are not taken into account
within this approach, is not to be described for the reaction
(p,xd).

The cross-section contribution of the multi-step compound process
for the (p,xp) reaction is by an order of magnitude smaller than
the multi-step direct mechanism.

The quantum-mechanical theory allows to play back the shape of spectra and
of the absolute cross-section for the reaction (p,xp). At the same time,
however, this program does not calculate (p,xd) reaction cross-sections
using the quantum-mechanical MSD and MSC model. We can use it only in the
calculations of Hauser-Feshbach components.

\begin{figure}
\includegraphics[width=14.0cm,height=20.5cm]{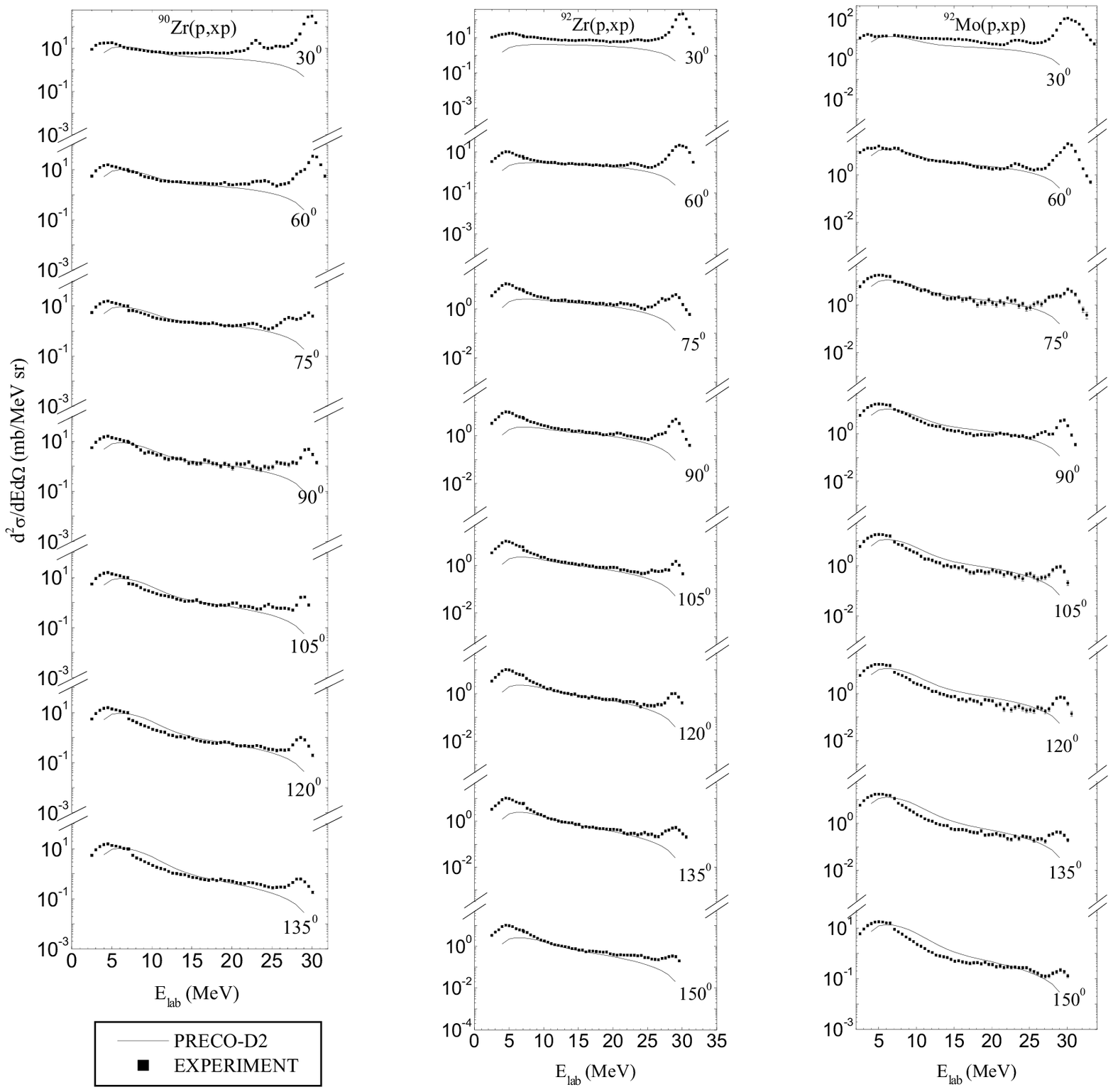}
\caption{Double-differential cross sections of reactions
$^{90}$Zr, $^{92}$Zr and $^{92}$Mo(p,xp)} \label{fig3}
\end{figure}

\begin{figure}
\includegraphics[width=14.0cm,height=20.5cm]{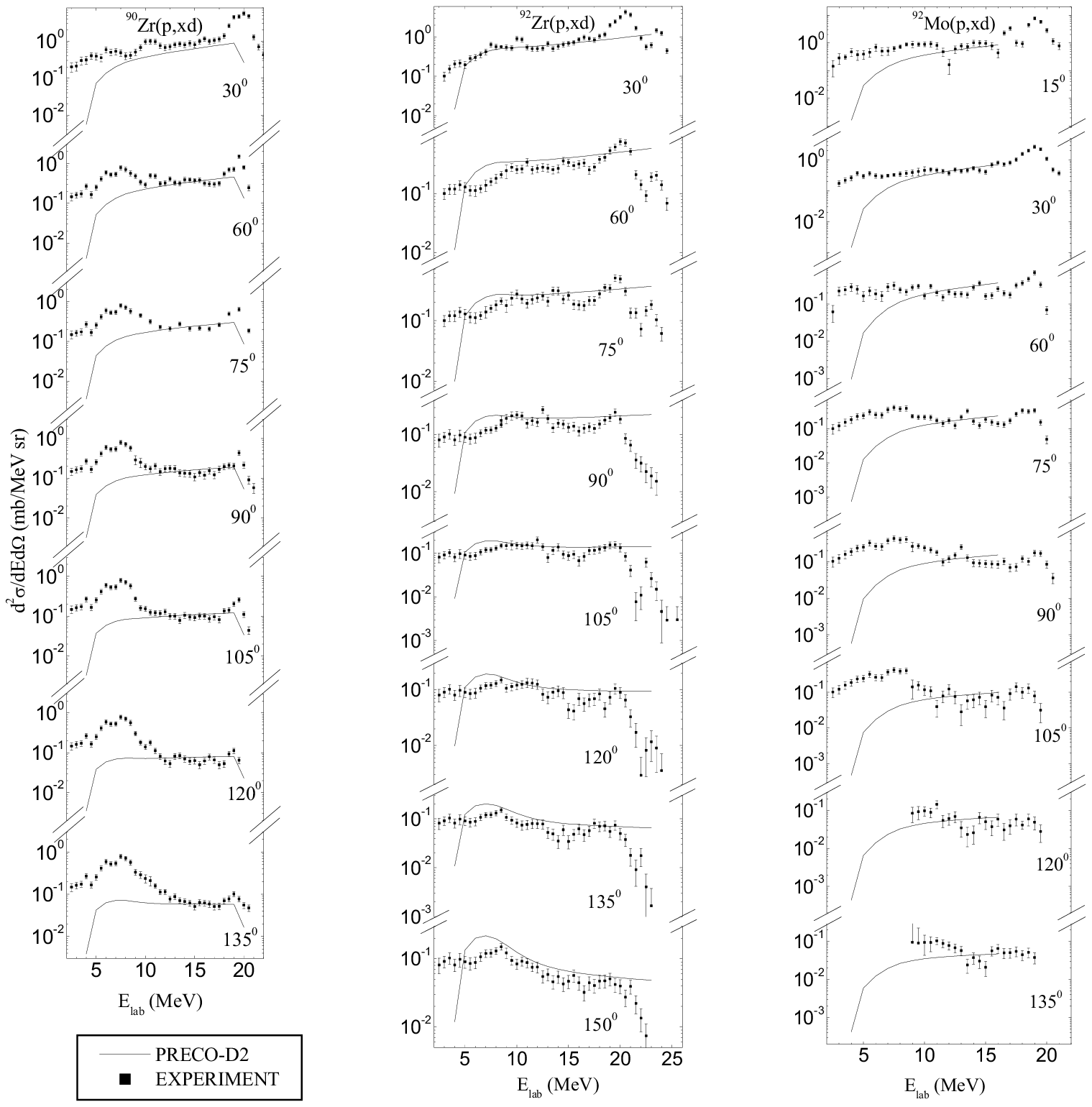}
\caption{Double-differential cross sections of reactions
$^{90}$Zr, $^{92}$Zr and $^{92}$Mo(p,xd)} \label{fig4}
\end{figure}

\begin{figure}
\includegraphics[width=14.0cm,height=20.5cm]{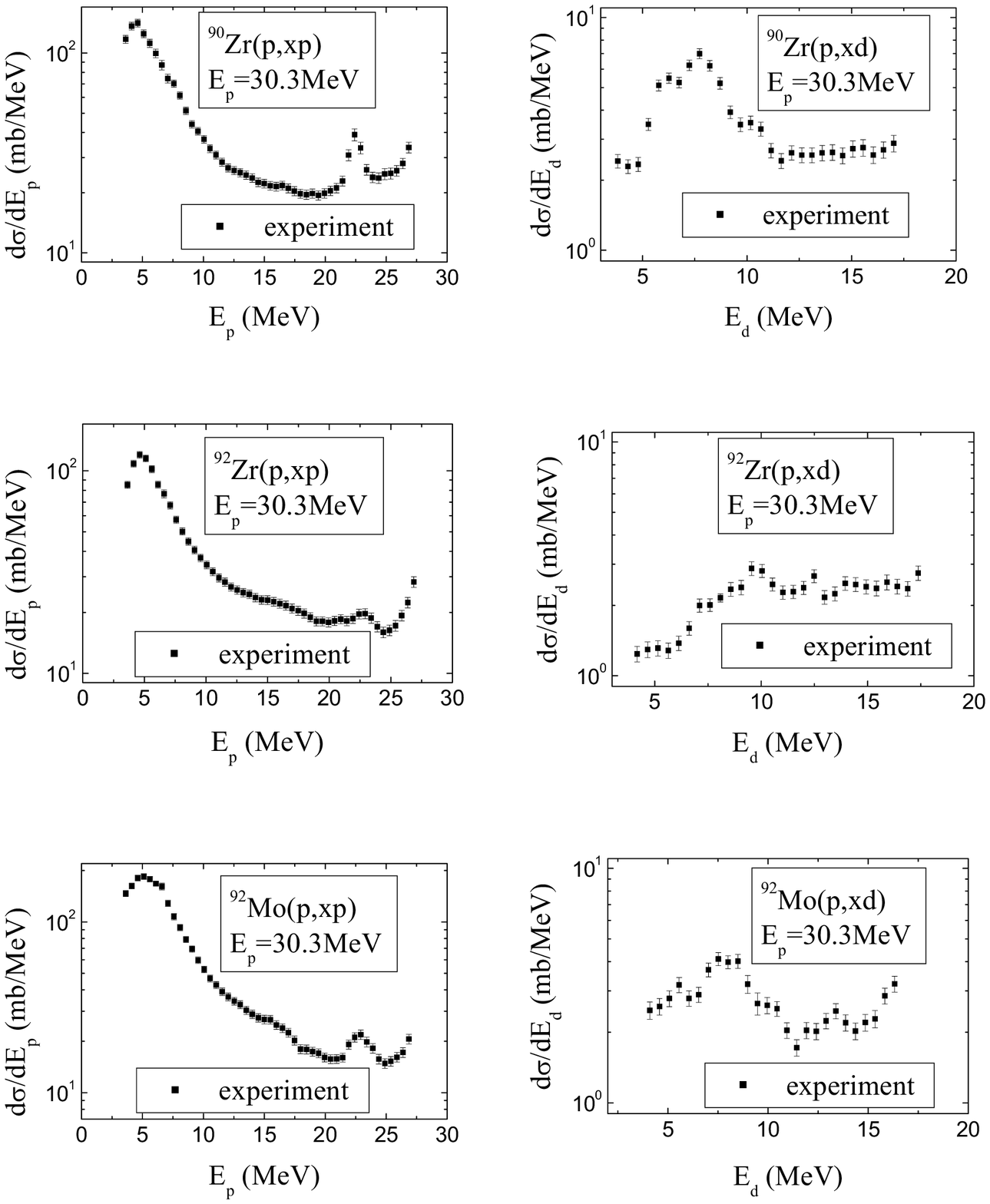}
\caption{Integral cross-sections of reactions $^{90}$Zr,
$^{92}$Zr, $^{92}$Mo(p,xp),(p,xd)} \label{fig5}
\end{figure}

\begin{figure}
\includegraphics[width=14.0cm,height=20.5cm]{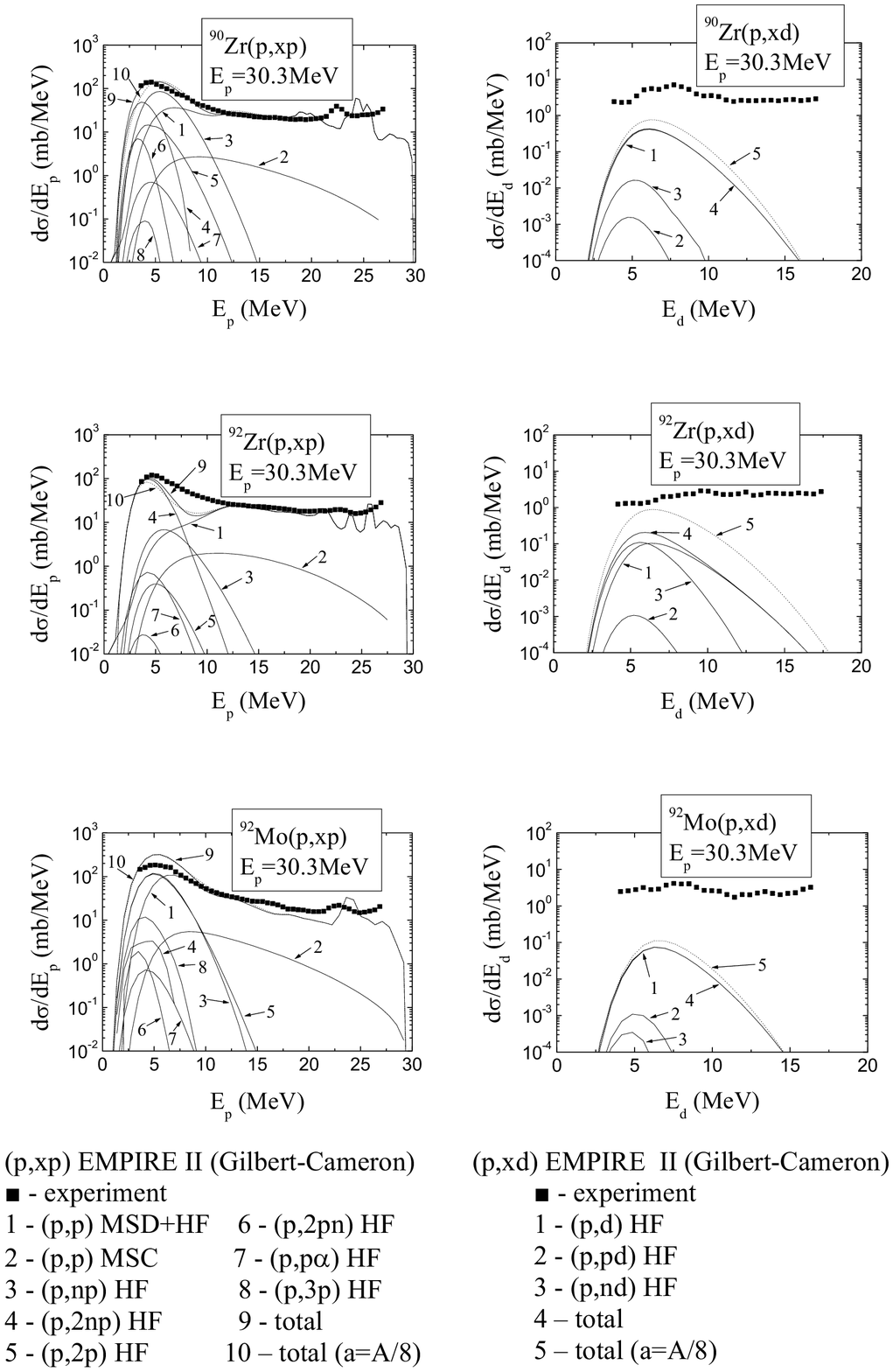}
\caption{Contribution MSC and MSD mechanisms to the formation of
integral cross-sections of reactions (p,xp) and (p,xd) obtained by
using the code EMPIRE.}  \label{fig6}
\end{figure}

\begin{figure}
\includegraphics[width=14.0cm,height=20.5cm]{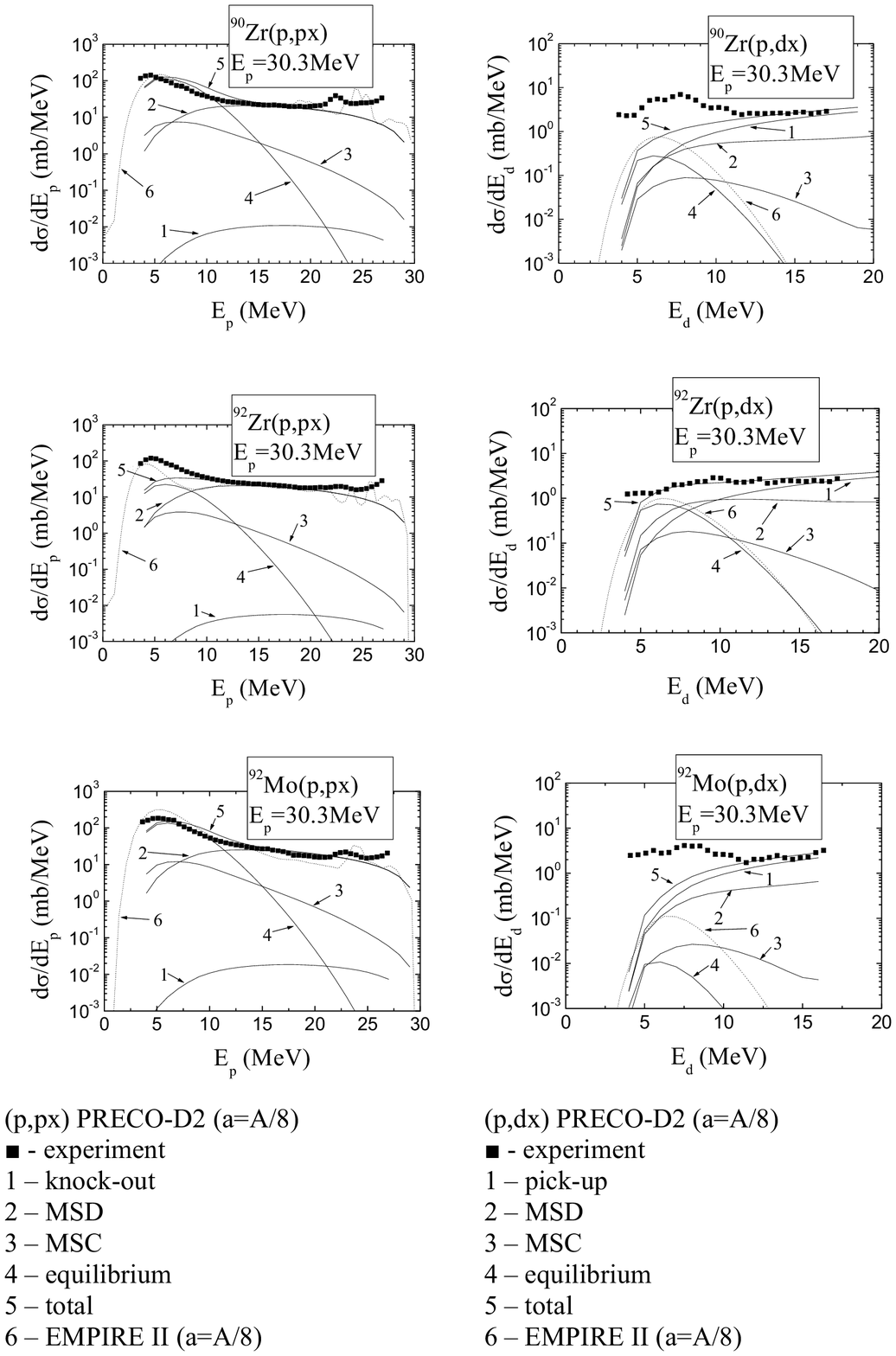}
\caption{Contribution of MSC and MSD mechanisms in the formation
of integral cross-sections of reactions (p,xp) and (p,xd) obtained
by using the code PRECO-D2.}  \label{fig7}
\end{figure}

In this connection, the calculations of the double differential
cross-sections (Figure \ref{fig3}, \ref{fig4}) and integral
spectra (Figure \ref{fig7}) have been carried out by using the
program PRECO-D2 \cite{ref9}. This code uses the Griffin exciton
model \cite{ref10} for pre-equilibrium nuclear reactions to
describe the emission of particles with mass numbers of 1 to 4
from an equilibrating composite nucleus. A distinction is made
between open and closed configurations in this system and between
the multi-step direct and multi-step compound components of the
pre-equilibrium cross sections \cite{ref13}. Additional MSD
components are calculated semi-empirically to account for direct
nucleon transfer reactions and direct knock-out processes
involving cluster degrees of freedom. Evaporation from the
equilibrated composite nucleus is included in the full MSC cross
section. Output of energy differential and double differential
cross sections is provided for the first particle emitted from the
composite system. For that, there are additional subroutines in
this program, which  use the total MSD (including direct) and
total MSC (including evaporation) cross sections to calculate the
angular distributions for the emitted particles. This is done
phenomenologically according to the paper by Kalbach and Mann
\cite{ref14}.

The comparisons of the experimental integral cross-sections with
PRECO-D2 calculations are shown in Figure \ref{fig7}. In all
calculations, the configuration (1p0h) has been firstly used. The
Becchetti-Greenlees optical potential parameters for proton
channel have been used to generate reaction transmission
coefficients. The optical potential of Mani {\it et al.}
\cite{ref15} has been used in the calculations for neutron channel
and the potential of Cline \cite{ref16} for deuterons. The
densities of levels have been parameterized as a=A/8. With the
same parameterization, the total cross-section for reaction (p,xp)
and multi-particle component for reaction (p,xd), calculated by
the program EMPIRE, are demonstrated for a comparison in Figure
\ref{fig7}. Figure \ref{fig3},  \ref{fig4} and \ref{fig7} show
that the satisfactory theoretical description of the
double-differential and integral cross-sections both for reactions
(p,xp) and (p,xd) has been reached. The main difference between
PRECO-D2 calculations and the experiment data is observed in a
soft part of the spectra that is apparently connected with the
neglect of the multi-particle emission within the code PRECO-D2.
The high-energy part of experimental spectra consists of an
elastic peak that can not be studied within the framework of this
model. The contributions of the MSD, the MSC, the equilibrium
components and the single-step direct process of knock-out for the
reaction (p,xp) and one-nucleon transfer (pick-up of a neutron)
for the reaction (p,xd) have been well estimated.

Two classes of single-step direct reactions which are not included
in the Griffin model are nucleon transfer (pickup) and knock-out
or inelastic-scattering processes which involve complex particle
degrees of freedom. While taking into account the contribution of
these direct processes, the description of the measured
experimental data can be considerably improved by the program
PRECO-D2. It follows from the theoretical results that the main
contribution into the cross-sections of the reaction (p,xd) is the
introduction of the mechanism of one-nucleon transfer. At the same
time, the contribution of the knock-out mechanism to the reaction
(p,xp) is much less, but the contribution of the MSD dominates and
that is in agreement with the results of calculations in the
quantum-mechanical theory. Also, it is possible to see that the
relative contribution of the MSC mechanism calculated in the
exciton model, correctly replicates the contribution from the MSC,
within the program EMPIRE.

\section{SUMMARY AND CONCLUSIONS}
\label{conc}

We have measured the double-differential $^{90,92}$Zr,$^{
92}$Mo(p,xp),(p,xd) reaction cross-sections at E$_{p}$=30.3$\pm
$0.15 MeV in order to investigate pre-equilibrium nucleon
emission. Integral d$\sigma $/dE spectra and partial reaction
cross sections have also been deduced. The experimental data has
been analyzed within the framework of the phenomenological exciton
model of pre-equilibrium decay and microscopic theory of the MSD
and MSC processes. Since there are no experimental data for the
protons around 30 MeV, this experimental study is very important
for the extension of the pre-equilibrium experiments in this
direction to see the mechanism of the reaction and the level of
energy-dependence. It is also important to observe the adequacy of
the above-mentioned theoretical models to explain the measured
experimental data.

For this purpose, we have presented that the satisfactory
theoretical description of the double-differential and integral
cross-sections both for reactions (p,xp) and (p,xd) has been
reached. The main difference between the PRECO-D2 calculation and
the experiment is observed in a soft part of the spectra. This
might be due to the neglect of the multi-particle emission in the
calculations. However, the contributions of the MSD, the MSC, the
equilibrium components and the single-step direct process of
knock-out mechanisms for the reaction (p,xp) and one-nucleon
transfer (pick-up of a neutron) for the reaction (p,xd) have been
well estimated. We have shown that it is also possible to improve
the description of the measured experimental data considerably by
taking into account the contribution of these direct processes. It
follows from the theoretical results that the main contribution
into the cross-sections of the reaction (p,xd) introduces the
mechanism of one-nucleon transfer. At the same time, the
contribution of the knock mechanism to the reaction (p,xp) is much
less, but the contribution of the MSD dominates and that is in
agreement with the results of calculations in the
quantum-mechanical theory. It is now well established that the
contribution of the multi-particle compound emission into
cross-sections of reactions is determined by protons with energies
from 2.5 up to 10 MeV, and from 5 MeV up to the kinematic limit in
direct processes.
\begin{acknowledgments}
Authors wishes to thank Professors P. Hodgson, R.S. Mackintosh, N.
Burtebaev and Dr. N. A. Boztosun for useful comments and careful
reading of the manuscript. I. Boztosun is also grateful to the
members of the Nuclear Physics Laboratory in Oxford University.
\end{acknowledgments}
\end{document}